\documentclass[11pt,a4paper]{article}
\usepackage[english]{babel}
\usepackage{amsmath,amsthm,amssymb,epsfig,latexsym}
\usepackage{easybmat}
\usepackage{color}
\usepackage{ulem}
\usepackage[numbers,square,sort&compress]{natbib}
\usepackage{hyperref}

\newcommand{\so}{\scriptscriptstyle \rm I}
\newcommand{\st}{\scriptscriptstyle \rm I\hspace{-1pt}I}
\newcommand{\sth}{\scriptscriptstyle \rm I\hspace{-1pt}I\hspace{-1pt}I}
\newcommand{\stf}{\scriptscriptstyle \rm I\hspace{-1pt}V}

\newcommand{\bro}{\bar\rho}

\newcommand{\bu}{\bar u}
\newcommand{\bv}{\bar v}

\newcommand{\bw}{\bar w}

\newcommand{\oma}[2]{\omega^{(a)}_{#1;\ell}(#2)}
\newcommand{\omd}[2]{\omega^{(d)}_{#1;\ell}(#2)}

\newcommand{\be}[1]{\begin{equation}\label{#1}}
\newcommand{\ba}[1]{\begin{multline}\label{#1}}
\newcommand{\ee}{\end{equation}}
\newcommand{\ea}{\end{eqnarray}}

\newcommand{\tr}{\mathop{\rm tr}}

\newcommand{\Res}{\mathop{\rm Res}}

\newcommand{\dd}{\mathrm{d}}

\newcommand{\CC}{\mathbb{C}}

\newtheorem{prop}{Proposition}[section]

\newtheorem{cor}{Corollary}[section]

\def\qed{\hfill\nobreak\hbox{$\square$}\par\medbreak}
 \makeatletter
 \@addtoreset{equation}{section}
 \makeatother

\newcommand{\bea}{\begin{eqnarray}}
\newcommand{\eea}{\end{eqnarray}}

\begin{document}

\vspace{12pt}

\begin{center}
\begin{LARGE}
{\bf   Action of the monodromy matrix entries \\[2mm] in the generalized algebraic  Bethe ansatz}
\end{LARGE}

\vspace{40pt}

{\large G.~Kulkarni\footnote{giridhar.kulkarni@ens-lyon.fr}} \\

Univ Lyon, ENS de Lyon, Univ Claude Bernard Lyon 1, CNRS, Laboratoire de Physique, F-69342 Lyon, France\\

\vspace{40pt}

{\large N.~A.~Slavnov\footnote{nslavnov@mi-ras.ru}}\\
Steklov Mathematical Institute of Russian Academy of Sciences, Moscow, Russia\\

 \vspace{12mm}

\end{center}

\vspace{1cm}


\begin{abstract}
We consider an $XYZ$ spin chain within the framework of the generalized algebraic Bethe ansatz. We calculate the actions of monodromy  matrix elements
on Bethe vectors as a linear combination of new Bethe vectors. We also compute the multiple action of the gauge transformed monodromy matrix elements on the pre-Bethe vector and conceive the result in terms of a partition function of the 8-vertex model.
\end{abstract}

\vspace{4mm}

\textbf{Key words:} Generalized algebraic Bethe ansatz, Bethe vectors, gauge transformed mo\-no\-dromy matrix,  domain wall partition function.

\vspace{1cm}

\section{Introduction}

Quantum Inverse Scattering Method (QISM) developed by the Leningrad school \cite{FadST79,FT79,FadLH96} allows us to find the spectra of quantum integrable models
Hamiltonians. This method is also used to calculate correlation functions. A number of interesting results have been obtained in
this way in models with an $R$-matrix of the 6-vertex model
\cite{IzeK84,Kor87,KojKS97,JimMMN92,KitMT00,GohKS04,KitMST05,KitKMST09,KitKMST11,KitKMST12,CauHM05,PerSCHMWA06,PerSCHMWA07,CauCS07,BogIK93L,Sla22L}.

A completely anisotropic $XYZ$ Heisenberg magnet \cite{Hei28} also can be studied within the QISM framework. However, this model has an $R$-matrix of
the 8-vertex model \cite{Sut70,FanW70,Baxter71,Baxter-book}. This leads to the fact that the corresponding monodromy matrix does not have a vacuum vector. As a result, the algebraic Bethe ansatz is not applicable to the $XYZ$ chain in its traditional formulation and requires an essential generalization. A generalized algebraic Bethe ansatz applicable to XYZ model was formulated in \cite{FT79}. It allows us to obtain Bethe equations that determine the
spectrum of the Hamiltonian, as well as construct the eigenvectors of the
transfer matrix. The question arises of the applicability of this method to
the calculation of form factors and correlation functions.

The calculation of correlation functions within the QISM consists of several
stages. At the first stage, it is necessary to express local operators of the model under consideration
in terms of the monodromy matrix entries. It can be done by explicitly solving
the quantum inverse problem \cite{KitMT99}.  For the $XYZ$
chain, the quantum inverse problem was solved in \cite{GohK00} (see also \cite{MaiT00}).

At the next step, it is necessary to calculate the actions of the monodromy matrix  elements on the Bethe vectors. This task is very simple for models with the 6-vertex $R$-matrix. In fact, the method of algebraic Bethe ansatz just gives the result of the action of monodromy matrix  elements on the Bethe vector in the form of a linear combination of new Bethe vectors. In the case of the $XYZ$ chain, the situation is completely different, and the problem of calculating the action of the monodromy matrix  elements becomes extremely nontrivial. The present paper is devoted to this issue.

At the last step, one should calculate the arising scalar products of Bethe vectors. This problem was solved in \cite{S89} for models with the 6-vertex $R$-matrix.
For the $XYZ$ chain, this problem was partly solved in \cite{SlaZZ21} using the method developed in \cite{BS19}. We will show in this paper that the results obtained in \cite{SlaZZ21} are not sufficient for computing form factors.

In models with the 6-vertex $R$-matrix, Bethe vectors are constructed by applying the upper-right element of the monodromy matrix to the vacuum vector. As we have already noted, there is no such vector in the $XYZ$ chain. Therefore, within the framework of the generalized algebraic Bethe ansatz, one has to introduce a special gauge transformation of the monodromy matrix.
Successive action of the upper-right element of the gauge transformed monodromy matrix to some analogue of vacuum vector allows us to construct pre-Bethe vectors and then Bethe vectors as the Fourier transforms of the former.
As a result, it becomes a very difficult task to calculate the action of the original monodromy  matrix elements on such a vector.

Action formulas have another interesting feature. Under the action of several operators on the Bethe vector (multiple action formulas) in models with
a rational or trigonometric $R$-matrix, a partition function of the 6-vertex model with a domain wall boundary condition arises \cite{KitMT00,BelPRS13,PakRS13}. Therefore, it is interesting to compute multiple action formulas in the case of the 8-vertex $R$-matrix. Recall that the partition function of the
8-vertex model with the domain wall boundary condition was found in \cite{PRS08} using elliptic current algebras and in \cite{YZ09}
using algebraic Bethe ansatz for solid-on-solid (SOS) model. A representation for the partition function as a sum of determinants was found in
\cite{Ros09}.

The paper is organized as follows.  In section~\ref{S-N}, we give a brief description of the generalized algebraic Bethe ansatz. Here we introduce a gauge transformation of the monodromy matrix and construct the Bethe vectors. In section~\ref{S-AGTO}, we calculate the actions of the elements of the gauge transformed monodromy matrix on pre-Bethe vectors. The results obtained allow us to solve the main problem in section~\ref{SS-BV}: to calculate the actions of the original monodromy  elements on the Bethe vectors.
Finally,  in section~\ref{S-MAGTO}, we give as an example the multiple action for the upper diagonal element of the gauge transformed monodromy matrix.
We show that in the multiple action formula, a numerical coefficient $K_m$ arises with well known recursive properties and it can be seen as the partition
function of the 8-vertex model with domain wall boundary conditions. Moreover, for a particular case of free fermions
we manage to provide it a determinant representation for $K_m$ in section~\ref{SS-FF}.

At the end of this paper we have collected basic information about Jacobi theta-functions in  appendix~\ref{A-JTF} and give some cumbersome calculations in
appendices~\ref{A-R}.

\section{Generalized algebraic Bethe ansatz for the XYZ model\label{S-N}}

In this section, we provide basic information about the generalized algebraic Bethe ansatz.
The reader can get acquainted with this method in more detail in works \cite{FT79,SlaZZ21}.

The Hamiltonian of the $XYZ$ chain with periodic boundary condition is given by
\be{HamXYZ}
H=\sum_{j=1}^{N} \Bigl (J_x \sigma^x_j\sigma^x_{j+1}+
J_y \sigma^y_j\sigma^y_{j+1}+J_z \sigma^z_j\sigma^z_{j+1}\Bigr ),
\ee
where $J_{x,y,z}$ are real constants, and we assume that the number of sites $N$ is even. The Hamiltonian \eqref{HamXYZ} acts in
a Hilbert space $\mathcal{H}$ which is a tensor
product of local quantum spaces $\mathcal{H}=\mathcal{H}_1\otimes\mathcal{H}_2\otimes\cdots\otimes\mathcal{H}_N$.
Here each $\mathcal{H}_k\cong\mathbb{C}^2$. The spin operators $\sigma^{x,y,z}_k$ are Pauli matrices acting non-trivially in $\mathcal{H}_k$.

\subsection{$R$-matrix and monodromy matrix}

Within the QISM framework, the $XYZ$ spin chain is constructed by an
8-vertex $R$-matrix

\be{R-mat}
R(u)=\begin{pmatrix}
a(u)&0&0&d(u)\\
0&b(u)&c(u)&0\\
0&c(u)&b(u)&0\\
d(u)&0&0&a(u)\\
\end{pmatrix},
\ee
where
\be{abcd}
\begin{aligned}
& a(u)=\frac{2\theta_4(\eta |2\tau ) \, \theta_1(u+\eta |2\tau )\,
\theta_4(u|2\tau )}{\theta_2(0|\tau )\,\theta_4(0|2\tau )},
\\[8pt]
& b(u)=\frac{2\theta_4(\eta |2\tau ) \, \theta_4(u+\eta |2\tau )\,
\theta_1(u|2\tau )}{\theta_2(0|\tau )\,\theta_4(0|2\tau )},
\\[8pt]
& c(u)=\frac{2\theta_1(\eta |2\tau ) \, \theta_4(u+\eta |2\tau )\,
\theta_4(u|2\tau )}{\theta_2(0|\tau )\, \theta_4(0|2\tau )},
\\[8pt]
& d(u)=\frac{2\theta_1(\eta |2\tau ) \, \theta_1(u+\eta |2\tau )\,
\theta_1(u|2\tau )}{\theta_2(0|\tau )\, \theta_4(0|2\tau )}.
\end{aligned}
\ee
The definition of the Jacobi theta-functions is given in appendix~\ref{A-JTF}. The parameters $\eta$ and $\tau$ are related to the interaction constants $J_{x,y,z}$
of the Hamiltonian \eqref{HamXYZ} (see below).

The monodromy matrix of the $XYZ$ model is defined as a product of the $R$-matrices
\be{Monod-def}
\mathcal{T}(u)=R_{01}(u-\xi_1)R_{02}(u-\xi_2)\cdots R_{0N}(u-\xi_N),
\ee
where complex parameters $\xi_k$ are called inhomogeneities. Each $R$-matrix  $R_{0k}(u-\xi_k)$ in this formula acts in the tensor product
$\mathcal{H}_0\otimes \mathcal{H}_k$, where $\mathcal{H}_k$ is one of the local quantum spaces, and $\mathcal{H}_0\cong\mathbb{C}^2$ is called an
auxiliary space. Traditionally, the monodromy matrix is written as a $2\times2$ matrix in the auxiliary space $\mathcal{H}_0$
\be{Monod-def1}
\mathcal{T}(u)=\begin{pmatrix} A(u)& B(u)\\ C(u)& D(u)
\end{pmatrix},
\ee
where $A(u)$, $B(u)$, $C(u)$, and  $D(u)$ are operators acting in $\mathcal{H}$. The monodromy matrix \eqref{Monod-def1} satisfies an
$RTT$-relation
\be{RTT}
R_{12}(u-v)\mathcal{T}_1(u)\mathcal{T}_2(v)=\mathcal{T}_2(v) \mathcal{T}_1(u)R_{12}(u-v),
\ee
which holds in the tensor product $\mathbb{C}^2\otimes\mathbb{C}^2\otimes\mathcal{H}$. The subscripts in \eqref{RTT} show in which of the two auxiliary
spaces $\mathbb{C}^2$ the monodromy matrix $\mathcal{T}_k$ acts nontrivially. The relation \eqref{RTT} defines commutation relations between the
operators $A(u)$, $B(u)$, $C(u)$, and  $D(u)$.

A transfer matrix ${\sf T}(u)$ is the trace of the monodromy matrix with respect to the auxiliary space
\be{Transf-mat}
{\sf T}(u)={\tr}_0 \mathcal{T}(u)=A(u)+D(u).
\ee
It is the generating function of the integrals of motion.
The Hamiltonian \eqref{HamXYZ} arises  in the homogeneous limit when all $\xi_k=0$:
\be{Ham-TM}
\frac{\dd}{\dd u}\log {\sf T}(u)\Bigr|_{u=0}=
\frac{\theta_1'(0|\tau )}{2\theta_1(\eta |\tau )}\,
H +J_0 N \mathbf{1},
\ee
where $J_0=\frac{1}{2}\,
\theta_1'(\eta |\tau )/\theta_1(\eta |\tau )$, and $\mathbf{1}$ is the identity operator. Then the constants $J_{x,y,z}$ are
$$
J_x=\frac{\theta_4(\eta |\tau )}{\theta_4(0|\tau )}\,, \quad
J_y=\frac{\theta_3(\eta |\tau )}{\theta_3(0|\tau )}\,, \quad
J_z=\frac{\theta_2(\eta |\tau )}{\theta_2(0|\tau )}\,.
$$

Despite the fact that only a homogeneous case is needed to construct the Hamiltonian of the $XYZ$ chain, in what follows we will consider a more general
inhomogeneous model \eqref{Monod-def} with arbitrary complex inhomogeneities $\xi_k$. We emphasize, however, that we do this solely for reasons of generality.
In all the formulas below, the homogeneous limit is trivial.

\subsection{Gauge transformed monodromy matrix and vacuum}

In the  original formulation of the algebraic Bethe ansatz, we require the existence of a vacuum vector $|0\rangle\in\mathcal{H}$ such that it is
annihilated by the lower-left element of the monodromy matrix: $C(u)|0\rangle=0$,  $\forall u\in\mathbb{C}$. In the case of the 8-vertex
$R$-matrix, such the vacuum vector does not exist. Therefore, to construct Bethe vectors we need to introduce generalized gauge-transformed monodromy matrices. Let
\be{GaugeT}
\mathcal{T}_{k,l}(u)=M^{-1}_k(u) \mathcal{T}(u)M_l(u)=
\begin{pmatrix}
A_{k,l}(u)&B_{k,l}(u)
\\
C_{k,l}(u)& D_{k,l}(u)\end{pmatrix}.
\ee
Here
\be{Mku}
M_k(u)=\begin{pmatrix}
\theta_1(s_{k} +u|2\tau )&\gamma_k\theta_1(t_{k} -u|2\tau )
\\
\theta_4(s_{k} +u|2\tau )&\gamma_k\theta_4(t_{k} -u|2\tau )
\end{pmatrix},
\ee
where $s_k=s+k\eta$, $t_k=t+k\eta$,
$s,t\in \CC$ are arbitrary parameters and
\be{gamu}
\gamma_k=\frac2{\theta_2(x_k|\tau)\theta_2(0|\tau)}, \qquad\text{where}\qquad x_k=\frac{s_k+t_k}2.  
\ee
It is easy to check that
\be{detM}
\det M_k(u)= \frac{2\theta_1(y+u|\tau)}{\theta_2(0|\tau)}, \qquad\text{where}\qquad y=\frac{s-t}2.
\ee

With the gauge transformation, we are ultimately using the vertex-IRF transformation \cite{Baxter-book,Fel95,SlaZZ21} that relates the $R$-matrix of the 8-vertex model to the dynamical $R$-matrix of the 8vSOS model
\be{vIRF}
R_{12}(u-v)M_{1,\ell}(u)M_{2,\ell}(v+\sigma^z_1\eta)=
M_{1,\ell}(u+\sigma^z_2\eta)M_{2,\ell}(v)\bar R_{12}(u-v|x_\ell).
\ee
Here $\bar R_{12}(u|x)$ is the dynamical $R$-matrix
\be{Rsos}
\bar R_{12}(u|x)=
\begin{pmatrix}
    \bar a(u)  &  0 & 0  & 0    \\
    0&   \bar b^+(u)  & \bar c^+(u) & 0\\
    0&   \bar c^-(u)  & \bar b^-(u)    & 0\\
    0&  0 &  0 & \bar a(u)
\end{pmatrix},
\ee
where
\be{dyn_abc}
\begin{aligned}
& \bar a(u)=\theta_1(u+\eta|\tau),
\\[8pt]
& \bar b^\pm(u)=\frac{\theta_1(u|\tau)\theta_2(x\pm \eta|\tau)}{\theta_2(x|\tau)},
\\[8pt]
& \bar c^\pm(u)=\frac{\theta_1(\eta|\tau)\theta_2(x\pm u|\tau)}{\theta_2(x|\tau)}.
\end{aligned}
\ee
This allows one to construct a vacuum vector for gauge transformed monodromy matrices in a similar way to the algebraic Bethe ansatz.
Let us introduce a local vacuum vector $|\omega_k^l\rangle$ by
\be{vacloc}
|\omega_k^l\rangle=\begin{pmatrix}
\theta_1(s_{k+l-1}+\xi_k|2\tau )
\\
\theta_4(s_{k+l-1}+\xi_k|2\tau )
\end{pmatrix}
 \in \mathcal{H}_k.
\ee
The global vacuum vectors are defined as
\be{vacglob}
|\Omega ^l\rangle=|\omega_1^l\rangle\otimes |\omega_2^l\rangle\otimes \ldots
\otimes |\omega_N^l\rangle.
\ee
Then one can check that
\be{actvac}
\begin{aligned}
& C_{l,l+N}(u)|\Omega^l\rangle=0,
\\
& A_{l,l+N}(u)|\Omega^l\rangle=a(u)|\Omega^{l+1}\rangle,
\\
& D_{l,l+N}(u)|\Omega^l\rangle=d(u) |\Omega^{l-1}\rangle,
\end{aligned}
\ee
where
\be{ad}
a(u)=\prod_{k=1}^N \theta_1 (u - \xi_k  + \eta|\tau),  \qquad d(u)=\prod_{k=1}^N \theta_1 (u - \xi_k|\tau ).
\ee

The Bethe vectors are constructed by the successive action of the operators $B_{k,l}(u)$ on the global vacuum vector (see below).

\subsubsection{Commutation relations and Bethe vectors}

Before moving on, we introduce some new notation. In what follows, for brevity, we will omit the modular parameter in the notation of theta-functions
 if it is equal to $\tau$, namely, $\theta_a(\cdot)\equiv\theta_a(\cdot|\tau)$.

Let us also introduce three functions which will be often used below
\be{functions}
g(u,v)=\frac{\theta_1(\eta)}{\theta_1(u-v)},\qquad
f(u,v)=\frac{\theta_1(u-v+\eta)}{\theta_1(u-v)},\qquad
h(u,v)=\frac{\theta_1(u-v+\eta)}{\theta_1(\eta)}.
\ee

In what follows, we will constantly deal with sets of complex variables.
We  denote these sets by a bar: $\bu=\{u_1,\dots,u_n\}$, $\bv=\{v_1,\dots,v_m\}$ etc. As a rule, the number of elements in the
sets is not shown explicitly in the equations, however we give these cardinalities in
special comments to the formulas. We also introduce special subsets $\bu_j=\bu\setminus\{u_j\}$, $\bu_{j,k}=\bu\setminus\{u_j,u_k\}$ and so on.

In order to make the formulas more compact we use a shorthand notation for products of  functions \eqref{functions}.
 Namely, if the functions $g$, $f$, $h$ depend on a set (or two sets) of variables, this means that one should take the product over the corresponding set.
For example,
 \be{SH-prodllll}
g(v,\bu)=\prod_{u_l\in\bu} g(v,u_l),\quad
f(u_j,\bu_j)=\prod_{\substack{u_l\in\bu\\ l\ne j}} f(u_j,u_l),  \quad f(\bv,\bu)=\prod_{\substack{u_l\in\bu\\ v_k\in\bv}} f(v_k,u_l)\qquad\text{etc.}
 \ee
By definition, any product over the empty set is equal to $1$. A double product is equal to $1$ if at least one of the sets
is empty.

We also apply this convention to the products of the functions $a(u)$ and $d(u)$  \eqref{ad}
\be{ad1}
a(\bu)=\prod_{u_l\in\bu} a(u_l),  \qquad d(\bv)=\prod_{v_l\in\bv} d(v_l).
\ee

The $RTT$-relation \eqref{RTT} implies
certain commutation relations between the gauge transformed operators  $A_{k,l}(u)$, $B_{k,l}(u)$, $C_{k,l}(u)$, and  $D_{k,l}(u)$ \cite{SlaZZ21}. In order to obtain them more efficiently we can also rely on the vertex-IRF relation \eqref{vIRF}.
We will list here only few commutation relations that we need. First of all,
\be{BB}
A_{k+1,l+1}(u)A_{k,l}(v)=A_{k+1,l+1}(v)A_{k,l}(u), \quad B_{k,l}(u)B_{k-1,l+1}(v)=B_{k,l}(v)B_{k-1,l+1}(u),
\ee
\be{CC}
C_{k,l+1}(u)C_{k+1,l}(v)=C_{k,l+1}(v)C_{k+1,l}(u), \quad D_{k,l}(u)D_{k+1,l+1}(v)=D_{k,l}(v)D_{k+1,l+1}(u).
\ee
The second type of commutation relations is
\begin{multline}\label{AB}
A_{k,l}(u)B_{k-1,l+1}(v)=f(v,u)B_{k,l+2}(v)A_{k-1,l+1}(u)\\
+g(u,v)\frac{\theta_2(u-v+x_{l+1})}{\theta_2(x_{l+1})}B_{k,l+2}(u)A_{k-1,l+1}(v),
\end{multline}
and
\begin{multline}\label{DB}
D_{k,l}(u)B_{k-1,l+1}(v)=f(u,v)B_{k-2,l}(v)D_{k-1,l+1}(u)\\
+g(v,u)\frac{\theta_2(u-v+x_{k-1})}{\theta_2(x_{k-1})}B_{k-2,l}(u)D_{k-1,l+1}(v).
\end{multline}
Recall that $x=(s+t)/2$ and $x_p=x+p\eta$. These formulas are quite similar to the standard commutation relations of the algebraic Bethe ansatz.
Following the tradition, we call the first terms in the rhs of \eqref{AB} and  \eqref{DB} (the operators preserve their initial arguments) the first commutation scheme and the second terms (the operators exchange the arguments) the second commutation scheme.

Finally, we have the third type of commutation relation  between non-diagonal elements of gauged transformed monodromy matrices
\begin{multline}\label{CB}
	C_{\ell-r,\ell+r}(u)B_{\ell-r-1,\ell+r+1}(v)=
	\frac{\gamma_{\ell-r-1}^2}{\gamma_{\ell-r}\gamma_{\ell-r-2}}
	B_{\ell-r-2,\ell+r+2}(v)C_{\ell-r-1,\ell+r+1}(u)
	\\
	\hspace{30mm} +	g(u,v) \frac{\theta_2(u-v+x_{\ell+r+1})}{\theta_2(x_{\ell+r+1})}  A_{\ell-r-2,\ell+r}(v)D_{\ell-r-1,\ell+r+1}(u)
	\\
	- 	g(u,v) \frac{\theta_2(u-v+x_{\ell-r-1})}{\theta_2(x_{\ell-r-1})}   A_{\ell-r-2,\ell+r}(u)D_{\ell-r-1,\ell+r+1}(v).
\end{multline}

To construct eigenvectors of the transfer matrix we first define a pre-Bethe vectors as
\be{psin}
	|\psi_{n}^{\ell}(\bar u)\rangle=
	B_{\ell-1,\ell+1}(u_{n})B_{\ell-2,\ell+2}(u_{n-1})\cdots B_{\ell-n,\ell+n}(u_1)|\Omega^{l-n}\rangle,
\ee
where $\bu=\{u_1,\dots,u_{n}\}$ and $n=N/2$. Due to commutation relations \eqref{BB} this vector is symmetric over the set $\bu$. A Bethe vector is then
defined as a Fourier transform of the pre-Bethe vector
\be{BV}
|\hat\Psi^\nu_n(\bu)\rangle =\sum_{\ell\in\mathbb{Z}}e^{-i\pi\nu\eta\ell}|\psi_{n}^{\ell}(\bar u)\rangle.
\ee
If the parameters $\bu$ satisfy a system of Bethe equations, then $|\hat\Psi^\nu_n(\bu)\rangle$ becomes an eigenvector of the
transfer matrix ${\sf T}(u)$ \cite{FT79}.

For irrational values of $\eta$ the Fourier transform \eqref{BV} is rather
formal because convergence of the infinite series is problematic. Formal expressions
of this kind become really meaningful for rational $\eta$,
\be{eta1}
\eta =\frac{2P}{Q},
\ee
where $P,Q$ are mutually prime integers\footnote{%
A more general case when the Bethe vectors are well-defined is the case when $\eta$ is a point of finite order on the elliptic curve, i.e.,
$Q\eta = 2P_1+P_2\tau$ with some integer $Q$, $P_1$, $P_2$ \cite{FT79}. We restrict ourselves to real $\eta$ for simplicity.}. In this case all functions in question
become $Q$-periodic in $\ell$ and the infinite Fourier series \eqref{BV} can be
substituted by the finite sum
\be{BV1}
|\hat\Psi^\nu_n(\bu)\rangle =\sum_{\ell=0}^{Q-1} e^{-i\pi\nu\eta\ell}|\psi_{n}^{\ell}(\bar u)\rangle.
\ee

In what follows, we restrict ourselves to the case of rational $\eta$. It should be noted, however, that the formulas for the action of the monodromy matrix
entries on the generalized pre-Bethe vectors (see below) remain valid for arbitrary $\eta$.

\section{Actions of the gauge transformed operators\label{S-AGTO}}

We introduce a generalized pre-Bethe vector as
\be{psinr}
	|\psi_{n-r}^{\ell}(\bar u)\rangle=
	B_{\ell-r-1,\ell+r+1}(u_{n-r})B_{\ell-r-2,\ell+r+2}(u_{n-r-1})\cdots B_{\ell-n,\ell+n}(u_1)|\Omega^{l-n}\rangle,
\ee
where $\bu=\{u_1,\dots,u_{n-r}\}$ is a set of arbitrary complex numbers, and $r\in\mathbb{Z}$. This vector turns into usual $|\psi_{n}^{\ell}(\bar u)\rangle$ at $r=0$. A generalized
Bethe vector is then defined as
\be{BV2}
|\hat\Psi^\nu_{n-r}(\bu)\rangle =\sum_{\ell=0}^{Q-1} e^{-2\pi i\ell\eta}|\psi_{n-r}^{\ell}(\bar u)\rangle.
\ee
Strictly speaking,
such a vector can become an eigenvector of the $XYZ$ Hamiltonian only in those sectors where $r= 0\mod Q$. Generalised Bethe vectors from other sectors $r\neq 0\mod Q$, although not needed in the construction of the spectrum, are still accessible through the action of monodromy matrix elements and hence they are pertinent for our discussion.
To find this action, our first goal is to derive the actions of the gauge transformed operators $A_{\ell-r,\ell+r}$, $B_{\ell-r,\ell+r}$,
$C_{\ell-r,\ell+r}$, and $D_{\ell-r,\ell+r}$ on the generalized pre-Bethe vectors.

Let $b=n-r+1$. We also define a set $\bu=\{u_1,\dots,u_{n-r}, u_{n-r+1}\}$. Then $\bu_b=\{u_1,\dots,u_{n-r}\}$. The action of
the $B_{\ell-r,\ell+r}$ operator on the vector $|\psi^{\ell}_{n-r}(\bar u_b)\rangle$ follows directly from the definition of
the generalized pre-Bethe vectors
\be{singactB}
B_{\ell-r,\ell+r}(u_b)|\psi^{\ell}_{b-1}(\bar u_b)\rangle=|\psi^{\ell}_{b}(\bar u)\rangle.
\ee

\subsection{Actions of the $A_{\ell-r,\ell+r}$ and $D_{\ell-r,\ell+r}$ operators}

\begin{prop}\label{Prop-Act-AD}
The action of the operators $A_{\ell-r,\ell+r}(u_b)$ and $D_{\ell-r,\ell+r}(u_b)$ on the pre-Bethe vector $|\psi_{n-r}(\bar u_b)\rangle$ have the following form:
\be{singactA}
A_{\ell-r,\ell+r}(u_b)|\psi^{\ell}_{b-1}(\bar u_b)\rangle=\sum_{j=1}^b\frac{a(u_j)f(\bu_j,u_j)}
{h(u_b,u_j)}\frac{\theta_2(u_b-u_j+x_{\ell+r+1})}{\theta_2(x_{\ell+r+1})}
|\psi^{\ell+1}_{b-1}(\bu_j)\rangle,
\ee
%
\be{singactD}
D_{\ell-r,\ell+r}(u_b)|\psi^{\ell}_{b-1}(\bar u_b)\rangle=\sum_{j=1}^b\frac{d(u_j)f(u_j,\bu_j)}
{h(u_j,u_b)}\frac{\theta_2(u_b-u_j+x_{\ell-r-1})}{\theta_2(x_{\ell-r-1})}
|\psi^{\ell-1}_{b-1}(\bu_j)\rangle.
\ee
\end{prop}

In fact, formulas \eqref{singactA} and \eqref{singactD} were already obtained in \cite{FT79}. In order to verify this, we single out the term at
$j=b$, for example, in \eqref{singactA}. Then
\begin{multline}\label{singactA-1}
A_{\ell-r,\ell+r}(u_b)|\psi^{\ell}_{b-1}(\bar u_b)\rangle=a(u_b)f(\bu_b,u_b)|\psi^{\ell+1}_{b-1}(\bu_b)\rangle \\
+\sum_{j=1}^{b-1}a(u_j)f(\bu_{j,b},u_j)g(u_b,u_j)
\frac{\theta_2(u_b-u_j+x_{\ell+r+1})}{\theta_2(x_{\ell+r+1})}
|\psi^{\ell+1}_{b-1}(\bu_j)\rangle,
\end{multline}
where we used $f(u,v)/h(u,v)=g(u,v)$ and $h(u,u)=1$.

This formula can be derived via the standard arguments of the algebraic Bethe ansatz. Using the commutation relations
\eqref{AB}, we  move the operator $A_{\ell-r,\ell+r}$ to the right through the product of the operators $B_{\ell-r-k,\ell+r+k}$. Having reached the extreme right position, we obtain the operator $A_{\ell-n,\ell+n}(u_k)$, where $u_k$ is one of the elements of the set $\bu$. Acting on the vacuum $|\Omega^{l-n}\rangle$, this operator gives the function $a(u_k)$. Thus, we conclude that the general structure of the resulting expression has the form
\begin{equation}\label{singactA-gen}
A_{\ell-r,\ell+r}(u_b)|\psi^{\ell}_{b-1}(\bar u_b)\rangle=a(u_b)\Lambda_b|\psi^{\ell+1}_{b-1}(\bu_b)\rangle
+\sum_{j=1}^{b-1}a(u_j)\Lambda_j
|\psi^{\ell+1}_{b-1}(\bu_j)\rangle,
\end{equation}
where $\Lambda_j$, $j=1,\dots,b$, are numerical coefficients.

It is easy to see that to obtain the first term in the rhs of \eqref{singactA-gen} we should only  use the first commutation scheme of the commutation relation
\eqref{AB}. This immediately gives us $\Lambda_b=f(\bu_b,u_b)$.

To obtain explicit expressions for $\Lambda_j$ with $j<b$, it is enough to find $\Lambda_{b-1}$ due to the symmetry of $|\psi^{\ell}_{b-1}(\bar u_b)\rangle$ over $\bu_b$. Then permuting $A_{\ell-r,\ell+r}(u_b)$ and $B_{\ell-r-1,\ell+r+1}(u_{b-1})$ we must use the second commutation scheme, otherwise, we can not obtain the coefficient $a(u_{b-1})$ in the result. After this, we again should use the first commutation scheme when moving $A_{\ell-r-1,\ell+r+1}(u_{b-1})$ to the right position. This consideration gives us
$$
\Lambda_{b-1}=f(\bu_{b-1,b},u_{b-1})g(u_b,u_{b-1})
\frac{\theta_2(u_b-u_{b-1}+x_{\ell+r+1})}{\theta_2(x_{\ell+r+1})},
$$
leading to
$$
\Lambda_{j}=f(\bu_{j,b},u_{j})g(u_b,u_{j})
\frac{\theta_2(u_b-u_{j}+x_{\ell+r+1})}{\theta_2(x_{\ell+r+1})}.
$$
In this way, we reproduce equation \eqref{singactA-1}, which is equivalent to \eqref{singactA}. The action \eqref{singactD} can be proved similarly.

\subsection{Action of the $C_{\ell-r,\ell+r}$ operator}

\begin{prop}\label{Prop-Act-C}
The action of the operator $C_{\ell-r,\ell+r}(u_b)$ on the pre-Bethe vector $|\psi_{b-1}(\bar u_b)\rangle$ has the following form:
\begin{multline}\label{actC-formula}
	C_{\ell-r,\ell+r}(u_b)|\psi^{\ell}_{b-1}(\bar u_b)\rangle=
	\sum_{\underset{j\neq k}{j,k=1}}^{b}
	\bigg\lbrace
	a(u_j)d(u_k)
	\frac{f(\bar u_j,u_j)f(u_k,\bar u_k)}{f(u_k,u_j)}
	\\
	\times\frac{\theta_2(x_{\ell+r+1}+u_b-u_j)}{h(u_b,u_j)\theta_2(x_{\ell+r+1})}
	\frac{\theta_2(x_{\ell-r-1}+u_b-u_k)}{h(u_k,u_b)\theta_2(x_{\ell-r-1})}
	|\psi^{\ell}_{b-2}(\bar u_{j,k})\rangle
	\bigg\rbrace.
	\end{multline}
\end{prop}

\textsl{Proof.} The proof can be performed in the traditional algebraic Bethe ansatz manner. First of all, we find the coefficient of $a(u_j)d(u_k)$ for $j$ and $k$ fixed so that $j\ne k\ne b$. Using \eqref{BB} we reorder the arguments of the pre-Bethe vector as follows:
\be{reord1}
C_{\ell-r,\ell+r}(u_b)|\psi_{b-1}(\bar u_b)\rangle= C_{\ell-r,\ell+r}(u_b)
B_{\ell-r-1,\ell+r+1}(u_k)B_{\ell-r-2,\ell+r+2}(u_j)|\psi^{\ell}_{b-3}(\bar u_{b,j,k})\rangle,
\ee
where $\bar u_{b,j,k}=\bar u\setminus \{u_{b},u_j,u_k\}$. Permuting $C_{\ell-r,\ell+r}(u_b)$ and  $B_{\ell-r-1,\ell+r+1}(u_k)$ via \eqref{CB} we obtain three types of terms:
\be{3-types}
\begin{aligned}
& \Big( B_{\ell-r-2,\ell+r+2}(u_k)C_{\ell-r-1,\ell+r+1}(u_b)\Big)
B_{\ell-r-2,\ell+r+2}(u_j)|\psi^{\ell}_{b-3}(\bar u_{b,j,k})\rangle,\\
& \Big( A_{\ell-r-2,\ell+r}(u_k)D_{\ell-r-1,\ell+r+1}(u_b)\Big)
B_{\ell-r-2,\ell+r+2}(u_j)|\psi^{\ell}_{b-3}(\bar u_{b,j,k})\rangle,\\
& \Big( A_{\ell-r-2,\ell+r}(u_b)D_{\ell-r-1,\ell+r+1}(u_k)\Big)
B_{\ell-r-2,\ell+r+2}(u_j)|\psi^{\ell}_{b-3}(\bar u_{b,j,k})\rangle.
\end{aligned}
\ee
Here the numeric coefficients are omitted for brevity.

The first possibility in \eqref{3-types} does not suite us, since the final result will contain the operator $B_{\ell-r-2,\ell+r+2}(u_k)$. Thus, the resulting vector anyway depends on $u_k$. We also
should not use the second possibility, because we can not obtain $d(u_k)$ in this case. Thus, we should work only with the third type in \eqref{3-types}:
\begin{multline}\label{1-step}
C_{\ell-r,\ell+r}(u_b)|\psi^{\ell}_{b-1}(\bar u_b)\rangle=
-g(u_b,u_k)\frac{\theta_2(u_b-u_k+x_{\ell-r-1})}{\theta_2(x_{\ell-r-1})}A_{\ell-r-2,\ell+r}(u_b)D_{\ell-r-1,\ell+r+1}(u_k)\\
\times B_{\ell-r-2,\ell+r+2}(u_j)|\psi^{\ell}_{b-3}(\bar u_{b,j,k})\rangle+\mathcal{Z},
\end{multline}
where here and below $\mathcal{Z}$ means all the terms which do not contribute to the desired coefficient.

The action of $D_{\ell-r-1,\ell+r+1}(u_k)$ on $B_{\ell-r-2,\ell+r+2}(u_j)|\psi^{\ell}_{b-3}(\bar u_{b,j,k})\rangle$ should be direct (that is, we should use only the first
commutation scheme). Otherwise, we do not obtain $d(u_k)$ in the final result. Hence,
\begin{multline}\label{2-step}
C_{\ell-r,\ell+r}(u_b)|\psi^{\ell}_{b-1}(\bar u_b)\rangle=
-g(u_b,u_k)\frac{\theta_2(u_b-u_k+x_{\ell-r-1})}{\theta_2(x_{\ell-r-1})}A_{\ell-r-2,\ell+r}(u_b)\\
\times B_{\ell-r-3,\ell+r+1}(u_j)|\psi^{\ell-1}_{b-3}(\bar u_{b,j,k})\rangle+\mathcal{Z}.
\end{multline}
Permuting $A_{\ell-r-2,\ell+r}(u_b)$ and $B_{\ell-r-3,\ell+r+1}(u_j)$ we should use the second commutation scheme, otherwise, we can not obtain $a(u_j)$. After this, when acting with $A_{\ell-r-3,\ell+r+1}(u_j)$ on the vector  $|\psi^{\ell-1}_{b-3}(\bar u_{b,j,k})\rangle$ we should use only the first commutation scheme. Thus, we finally arrive at
\begin{multline}\label{3-step}
C_{\ell-r,\ell+r}(u_b)|\psi^{\ell}_{b-1}(\bar u_b)\rangle=
-g(u_b,u_k)g(u_b,u_j)\frac{\theta_2(u_b-u_k+x_{\ell-r-1})}{\theta_2(x_{\ell-r-1})}\frac{\theta_2(u_b-u_j+x_{\ell+r+1})}{\theta_2(x_{\ell+r+1})}
\\
\times d(u_k)a(u_j) f(u_k,\bu_{b,k})f(\bu_{b,j,k},u_j) |\psi^{\ell}_{b-3}(\bar u_{j,k})\rangle+\mathcal{Z}.
\end{multline}
It remains to check that
\be{check-1}
-g(u_b,u_k)g(u_b,u_j)f(u_k,\bu_{b,k})f(\bu_{b,j,k},u_j)=\frac{f(\bar u_j,u_j)f(u_k,\bar u_k)}{f(u_k,u_j)h(u_b,u_j)h(u_k,u_b)}, \qquad j\ne k\ne b.
\ee

We now consider the coefficient of $a(u_b)d(u_k)$ for $k \ne b$. We reorder the arguments of $|\psi_{b-1}(\bar u_b)\rangle$ as follows:
\be{reord0}
|\psi_{b-1}(\bar u_b)\rangle=
B_{\ell-r-1,\ell+r+1}(u_k)|\psi^{\ell}_{b-2}(\bar u_{b,k})\rangle.
\ee
Hence,
\be{reord2}
C_{\ell-r,\ell+r}(u_b)|\psi_{b-1}(\bar u_b)\rangle= C_{\ell-r,\ell+r}(u_b)
B_{\ell-r-1,\ell+r+1}(u_k)|\psi^{\ell}_{b-2}(\bar u_{b,k})\rangle.
\ee
Permuting $C_{\ell-r,\ell+r}(u_b)$ and $B_{\ell-r-1,\ell+r+1}(u_k)$ we again obtain three terms listed in \eqref{3-types}. And again,
only the third type in \eqref{3-types} can give us desired coefficient. We find
\begin{multline}\label{1-step0}
C_{\ell-r,\ell+r}(u_b)|\psi^{\ell}_{b-1}(\bar u_b)\rangle=
-g(u_b,u_k)\frac{\theta_2(u_b-u_k+x_{\ell-r-1})}{\theta_2(x_{\ell-r-1})}A_{\ell-r-2,\ell+r}(u_b)\\
\times D_{\ell-r-1,\ell+r+1}(u_k)|\psi^{\ell}_{b-2}(\bar u_{b,k})\rangle+\mathcal{Z}.
\end{multline}
Obviously, the actions of $D_{\ell-r-1,\ell+r+1}(u_k)$ and $A_{\ell-r-2,\ell+r}(u_b)$ should be direct (that is, we should use only the first commutation scheme in both cases).
We arrive at
\begin{multline}\label{2-step-2}
	C_{\ell-r,\ell+r}(u_b)|\psi^{\ell}_{b-1}(\bar u_b)\rangle=
	-a(u_b)d(u_k)g(u_b,u_k)\frac{\theta_2(u_b-u_k+x_{\ell-r-1})}{\theta_2(x_{\ell-r-1})}\\
\times f(\bar u_{b,k},u_b)f(u_k,\bar u_{b,k})	
	|\psi_{b-2}^{\ell}(\bar u_{b,k})\rangle +\mathcal{Z}.
	\end{multline}
We see that we reproduce the coefficient of $a(u_b)d(u_k)$ in \eqref{actC-formula}.

It remains to find the coefficient of $a(u_j)d(u_b)$ for $j\ne b$. This case is very similar to the previous one. After appropriate reordering of the arguments of
$|\psi^{\ell}_{b-1}(\bar u_b)\rangle$, we obtain
\be{reord3}
C_{\ell-r,\ell+r}(u_b)|\psi_{b-1}(\bar u_b)\rangle= C_{\ell-r,\ell+r}(u_b)
B_{\ell-r-1,\ell+r+1}(u_j)|\psi^{\ell}_{b-2}(\bar u_{b,j})\rangle.
\ee
After permutation of $C_{\ell-r,\ell+r}(u_b)$ and $B_{\ell-r-1,\ell+r+1}(u_j)$ we obtain three terms \eqref{3-types}, in which we should replace $j\leftrightarrow k$. Only one of these three terms (the second) gives the desired contribution:
\begin{multline}\label{2-step0}
C_{\ell-r,\ell+r}(u_b)|\psi^{\ell}_{b-1}(\bar u_b)\rangle=
g(u_b,u_j)\frac{\theta_2(u_b-u_j+x_{\ell+r+1})}{\theta_2(x_{\ell+r+1})}A_{\ell-r-2,\ell+r}(u_j)\\
\times D_{\ell-r-1,\ell+r+1}(u_b)|\psi^{\ell}_{b-2}(\bar u_{b,j})\rangle+\mathcal{Z}.
\end{multline}
Obviously, the actions of $D_{\ell-r-1,\ell+r+1}(u_b)$ and $A_{\ell-r-2,\ell+r}(u_j)$ should be direct, leading to	
\begin{equation}\label{2-step1}
C_{\ell-r,\ell+r}(u_b)|\psi^{\ell}_{b-1}(\bar u_b)\rangle=
g(u_b,u_j)\frac{\theta_2(u_b-u_j+x_{\ell+r+1})}{\theta_2(x_{\ell+r+1})}f(u_b,\bu_{b,j})f(\bu_{b,j},u_j)|\psi^{\ell}_{b-2}(\bar u_{b,j})\rangle+\mathcal{Z}.
\end{equation}
We  reproduce the coefficient of $a(u_j)d(u_b)$ in \eqref{actC-formula}. \qed

We conclude this section by writing down the action formulas in a compact way.
Let
\be{shrt}
\begin{aligned}
&\oma{j}{z}=\frac{a(u_j)f(\bu_j,u_j)}
{h(z,u_j)}\frac{\theta_2(z-u_j+x_{\ell+r+1})}{\theta_2(x_{\ell+r+1})},\\
&\omd{j}{z}=\frac{d(u_j)f(u_j,\bu_j)}
{h(u_j,z)}\frac{\theta_2(z-u_j+x_{\ell-r-1})}{\theta_2(x_{\ell-r-1})}.
\end{aligned}
\ee
Then the action formulas \eqref{singactA} and \eqref{singactD} take the form
\be{shrtAD}
\begin{aligned}
&A_{\ell-r,\ell+r}(u_b)|\psi^{\ell}_{b-1}(\bar u_b)\rangle=\sum_{j=1}^b \oma{j}{u_b}|\psi^{\ell+1}_{b-1}(\bar u_j)\rangle,\\
&D_{\ell-r,\ell+r}(u_b)|\psi^{\ell}_{b-1}(\bar u_b)\rangle=\sum_{j=1}^b \omd{j}{u_b}|\psi^{\ell-1}_{b-1}(\bar u_j)\rangle.
\end{aligned}
\ee
The action formula \eqref{actC-formula}  can be written as follows:
\be{shrtC}
C_{\ell-r,\ell+r}(u_b)|\psi^{\ell}_{b-1}(\bar u_b)\rangle=
	\sum_{\substack{j,k=1\\ j\neq k}}^{b}
	\frac{\oma{j}{u_b}\omd{k}{u_b}}{f(u_k,u_j)} 	|\psi^{\ell}_{b-2}(\bar u_{j,k})\rangle .
\ee

\section{Actions of the monodromy matrix entries on Bethe vectors\label{SS-BV}}

Now we can easily find the actions of the operators of the original monodromy matrix on the generalized pre-Bethe vectors. For this, it enough to express
these operators through the gauge transformed ones using formula \eqref{GaugeT}. It is also convenient to pass from the original operators to the operators
$A(u)\pm D(u)$ and $B(u)\pm C(u)$, since the local spin operators $\sigma_k^{x,y,z}$ are expressed precisely in terms of such combinations in
the framework of the quantum inverse problem \cite{GohK00}.

Let us introduce two $4$-components vectors consisting of the monodromy matrix entries
\be{VectTT}
\mathbf{T}(u)=\begin{pmatrix}
A(u)+D(u)\\
A(u)-D(u)\\
B(u)-C(u)\\
B(u)+C(u)
\end{pmatrix}, \qquad\qquad
\mathbf{T}^{(\ell,r)}(u)=
\begin{pmatrix}
A_{\ell-r,\ell+r}(u)\\
D_{\ell-r,\ell+r}(u)\\
C_{\ell-r,\ell+r}(u)\\
B_{\ell-r,\ell+r}(u)
\end{pmatrix}.
\ee
Then it follows from \eqref{GaugeT} that
\be{VectAct}
\mathbf{T}(u)=\frac{\mathbf{W}^{(\ell,r)}(u)}{\theta_1(y+u)}\mathbf{T}^{(\ell,r)}(u).
\ee
A $4\times4$ matrix $\mathbf{W}^{(\ell,r)}(u)$ has the following entries:
\be{Wmat0}
\mathbf{W}^{(\ell,r)}(u)=\left(
\begin{array}{cccc}
\frac{\theta_1(y_{-r}+u)\theta_2(x_\ell)}{\theta_2(x_{\ell+r})}&\frac{ \theta_1(y_{r}+u)\theta_2(x_\ell)}{\theta_2(x_{\ell-r})}
&\frac{ -2\theta_1(r\eta)\theta_2(t_\ell-u) }{\theta_2(0)\theta_2(x_{\ell-r})\theta_2(x_{\ell+r})}&\frac{\theta_2(0)\theta_1(r\eta)\theta_2(s_\ell+u)}{2}\\[8pt]
\frac{\theta_2(y_{-r}+u)\theta_1(x_\ell)}{\theta_2(x_{\ell+r})}&\frac{ -\theta_2(y_{r}+u)\theta_1(x_\ell)}{\theta_2(x_{\ell-r})}
&\frac{ 2\theta_2(r\eta)\theta_1(t_\ell-u) }{\theta_2(0)\theta_2(x_{\ell-r})\theta_2(x_{\ell+r})}&\frac{-\theta_2(0)\theta_2(r\eta)\theta_1(s_\ell+u)}{2}\\[8pt]
\frac{-\theta_3(y_{-r}+u)\theta_4(x_\ell)}{\theta_2(x_{\ell+r})}&\frac{ \theta_3(y_{r}+u)\theta_4(x_\ell)}{\theta_2(x_{\ell-r})}
&\frac{-2\theta_3(r\eta)\theta_4(t_\ell-u) }{\theta_2(0)\theta_2(x_{\ell-r})\theta_2(x_{\ell+r})}&\frac{\theta_2(0)\theta_3(r\eta)\theta_4(s_\ell+u)}{2}\\[8pt]
\frac{\theta_4(y_{-r}+u)\theta_3(x_\ell)}{\theta_2(x_{\ell+r})}&\frac{-\theta_4(y_{r}+u)\theta_3(x_\ell)}{\theta_2(x_{\ell-r})}
&\frac{ 2\theta_4(r\eta)\theta_3(t_\ell-u) }{\theta_2(0)\theta_2(x_{\ell-r})\theta_2(x_{\ell+r})}&\frac{-\theta_2(0)\theta_4(r\eta)\theta_3(s_\ell+u)}{2}
\end{array}\right),
\ee
where $y_{\pm r}=y\pm r\eta$.

Using \eqref{singactB}, \eqref{shrtAD}, and \eqref{shrtC} we immediately obtain
\begin{multline}\label{actABCD}
\mathbf{T}_p(u_b)|\psi^{\ell}_{b-1}(\bar u_b)\rangle\\
= \frac{1}{\theta_1(y+u_b)}\sum_{j=1}^b \Big[
\mathbf{W}^{(\ell,r)}_{p1}(u_b)\oma{j}{u_b}|\psi^{\ell+1}_{b-1}(\bar u_j)\rangle
+\mathbf{W}^{(\ell,r)}_{p2}(u_b)\omd{j}{u_b}|\psi^{\ell-1}_{b-1}(\bar u_j)\rangle\Big]\\
+\sum_{\substack{j,k=1\\ j\neq k}}^{b}\mathbf{W}^{(\ell,r)}_{p3}(u_b)
	\frac{\oma{j}{u_b}\omd{k}{u_b}}{f(u_k,u_j)} 	|\psi^{\ell}_{b-2}(\bar u_{j,k})\rangle
+\mathbf{W}^{(\ell,r)}_{p4}(u_b)|\psi^{\ell}_{b}(\bar u)\rangle,
\end{multline}
for $p=1,2,3,4$.

To obtain action formulas of the operators $\mathbf{T}_p(u_b)$ on the Bethe vectors we need to take the Fourier transform of \eqref{actABCD}.
Let us introduce the following Fourier transforms:
\be{FourierW}
\begin{aligned}
&\widehat{\mathbf{W}}^{(\nu)}_{p1}(u_b,u_j)=\sum_{\ell=0}^{Q-1}e^{-i\pi\nu\eta\ell}\;\mathbf{W}^{(\ell,r)}_{p1}(u_b)\oma{j}{u_b},\\
&\widehat{\mathbf{W}}^{(\nu)}_{p2}(u_b,u_j)=\sum_{\ell=0}^{Q-1}e^{-i\pi\nu\eta\ell}\;\mathbf{W}^{(\ell,r)}_{p2}(u_b)\omd{j}{u_b},\\
&\widehat{\mathbf{W}}^{(\nu)}_{p3}(u_b,u_j,u_k)=\sum_{\ell=0}^{Q-1}e^{-i\pi\nu\eta\ell}\;\mathbf{W}^{(\ell,r)}_{p3}(u_b)\oma{j}{u_b}\omd{k}{u_b},\\
&\widehat{\mathbf{W}}^{(\nu)}_{p4}(u_b)=\sum_{\ell=0}^{Q-1}e^{-i\pi\nu\eta\ell}\;\mathbf{W}^{(\ell,r)}_{p4}(u_b).
\end{aligned}
\ee
Recall that here we consider $\eta=2P/Q$. At the same time, equation \eqref{actABCD} holds for arbitrary complex $\eta$.
Using the fact that Fourier transform of a product gives a convolution of the Fourier transforms we obtain
\begin{multline}\label{actABCD-BV}
\mathbf{T}_p(u_b)|\hat\Psi^{\nu}_{b-1}(\bar u_b)\rangle\\
=\frac1{Q\theta_1(y+u_b)}\sum_{\mu=0}^{Q-1}\Bigg\{\sum_{j=1}^b \Big[
e^{i\pi\eta\mu}\widehat{\mathbf{W}}^{(\nu-\mu)}_{p1}(u_b,u_j)
+e^{-i\pi\eta\mu}\widehat{\mathbf{W}}^{(\nu-\mu)}_{p2}(u_b,u_j)\Big]|\hat\Psi^{\mu}_{b-1}(\bar u_j)\rangle\\
+\sum_{\substack{j,k=1\\ j\neq k}}^{b}
	\frac{\widehat{\mathbf{W}}^{(\nu-\mu)}_{p3}(u_b,u_j,u_k)}{f(u_k,u_j)} 	|\hat\Psi^{\mu}_{b-2}(\bar u_{j,k})\rangle
+\widehat{\mathbf{W}}^{(\nu-\mu)}_{p4}(u_b)|\hat\Psi^{\mu}_{b}(\bar u)\rangle\Bigg\}.
\end{multline}
Thus, acting with the operators $\mathbf{T}_p(u_b)$ on the Bethe vector $|\hat\Psi^{\nu}_{b-1}(\bar u_b)\rangle$ we obtain Bethe vectors of
three types: $|\hat\Psi^{\mu}_{b}(\bar u)\rangle$, $|\hat\Psi^{\mu}_{b-1}(\bar u_j)\rangle$, and $|\hat\Psi^{\mu}_{b-2}(\bar u_{j,k})\rangle$.

\section{Multiple action of the gauge transformed operators\label{S-MAGTO}}

We have mentioned already that in models with rational and trigonometric $R$-matrices, it is possible to calculate the actions of not only one operator on the Bethe vector, but also the actions of the product of several operators \cite{KitMT00,BelPRS13,PakRS13,HutLPRS16}. We call such actions multiple actions.
We consider the analogous case for the multiple actions of  gauged transformed monodromy matrix elements on the pre-Bethe vectors.

As an example, let us consider the action of a product of the gauge transformed operators $A_{\ell+k-r,\ell+k+r}(v_{k+1})$.
Let
\be{bbA}
\mathbb{A}^\ell_{m,n-r}(\bv)= A_{\ell+m-1-r,\ell+m-1+r}(v_m)\cdots A_{\ell+1-r,\ell+1+r}(v_2)A_{\ell-r,\ell+r}(v_1).
\ee
Note that the operator $\mathbb{A}^\ell_{m,n-r}(\bv)$ is symmetric over $\bv=\{v_1,\dots,v_{m}\}$ due to commutation relations \eqref{BB}.
Consider the action of $\mathbb{A}^\ell_{m,n-r}(\bv)$ on the vector $|\psi^{\ell}_{n-r}(\bar u)\rangle$, where $\bu=\{u_1,\dots,u_{n-r}\}$. It is clear that  the result of this action can be written in the following form
\be{singactA-m}
\mathbb{A}^\ell_{m,n-r}(\bv)|\psi^{\ell}_{n-r}(\bar u)\rangle=\sum_{\{\bro_{\so},\bro_{\st}\}\vdash\{\bv,\bu\}}\Lambda_m^{(\ell,r)}(\bro_{\so},\bro_{\st})|\psi^{\ell+m}_{n-r}(\bro_{\st})\rangle.
\ee
Here the sum is taken over partitions of the union $\{\bv,\bu\}\equiv\bro$ into two subsets $\bro_{\so}$ and $\bro_{\st}$ so that $\#\bro_{\so}=m$,
and $\Lambda^{(\ell,r)}(\bro_{\so},\bro_{\st})$ are numerical coefficients to be determined.
Indeed, the successive action of the operators $A_{\ell+k-r,\ell+k+r}(v_{k+1})$ on the vector $|\psi^{\ell}_{n-r}(\bar u)\rangle$ gives a linear combination of vectors $|\psi^{\ell+m}_{n-r}\rangle$ depending on all possible subsets of $\{\bv,\bu\}$,
consisting of $n-r$ elements. Equation \eqref{singactA-m} is the most general formula of this kind.

We will look for the coefficients $\Lambda^{(\ell,r)}(\bro_{\so},\bro_{\st})$ in the form
\be{Lam-K}
\Lambda_m^{(\ell,r)}(\bro_{\so},\bro_{\st})=\frac{a(\bro_{\so})K_m^{ p}(\bv|\bro_{\so})}
{f(\bv,\bro_{\so})}f(\bro_{\st},\bro_{\so}),
\ee
where $K_m^{{p}}(\bv|\bro_{\so})$ is a new unknown function to be determined. {All gauge dependent coefficients are contained in this term. Moreover, we can also see from \eqref{singactA-1}  that it only depends on the sum $p=\ell+r$, hence the choice of our notation. The remaining gauge independent terms contain products over known functions. Let us recall that we use conventions \eqref{SH-prodllll} and \eqref{ad1} for writing these terms}.

\textsl{Remark}. We use ansatz \eqref{Lam-K} by analogy with multiple action formulas in models with the 6-vertex $R$-matrix.
In this case, the multiple action of the
operators $A(v)$ on the Bethe vector is given by formulas \eqref{singactA-m} and \eqref{Lam-K}, and the coefficient $K_m(\bv|\bro_{\so})$ is the partition function of the 6-vertex model with domain wall boundary conditions (see \cite{BelPRS13,PakRS13}), where the latter admits a determinant representation \cite{I87,K82} known as the Izergin--Korepin formula.

Setting $m=1$ in \eqref{singactA-m}, \eqref{Lam-K} and comparing these equations with \eqref{singactA} we obtain
\begin{equation}\label{rec-002}
K_{1}^{{p}}(v|w)= g(v,w) \frac{\theta_2(v-w+x_{p+1})}{\theta_2(x_{p+1})}.
\end{equation}
The function $K_m^{{p}}(\bv|\bw)$ with $m>1$ can be obtained recursively due to the following proposition.

\begin{prop}\label{Prop-Km}
The function $K_m^{{p}}(\bv|\bw)$ satisfies the following identity:
\begin{equation}\label{recur-2}
K_{m}^{p}(\bv|\bw)= \sum_{\{\bw_{\so},\bw_{\st}\}\vdash \bw} K^p_{m_1}(\bv_{\so}|\bw_{\so})K^{p +m_1}_{m-m_1}(\bv_{\st}|\bw_{\st})
f(\bw_{\st},\bw_{\so}) f(\bv_{\so},\bw_{\st}).
\end{equation}
Here $1<m_1<m$, and $\bv_{\so}$ and $\bv_{\st}$ are arbitrary fixed subsets of $\bv$ with cardinalities $m_1$ and $m-m_1$ respectively. The sum is taken over partitions of the set $\bw$ into subsets  $\bw_{\so}$ and $\bw_{\st}$ such that $\#\bw_{\so}=m_1$ and $\#\bw_{\st}=m-m_1$.
\end{prop}

The proof of this proposition is given in appendix~\ref{A-R}.

\textsl{Remark.}
Setting $K_{0}^{{p}}(\emptyset|\emptyset)=1$ by definition, we extend the statement of proposition~\ref{Prop-Km} to the cases $m_1=0$ and $m_1=m$.

\begin{cor}\label{Cor-Km}
The function $K_m^{{p}}(\bv|\bw)$ satisfies the following recursions
\begin{equation}\label{rec-001}
K_{m}^{{p}}(\bv|\bw)= \sum_{k=1}^m g(v_m,w_k) \frac{\theta_2(v_m-w_k+x_{p+m})}{\theta_2(x_{p+m})}f(w_k,\bw_k)
f(\bv_m,w_k)K_{m-1}^{{p}}(\bv_m|\bw_k),
\end{equation}
and
\begin{equation}\label{rec-001a}
K_{m}^{p}(\bv|\bw)= \sum_{k=1}^m g(v_m,w_k) \frac{\theta_2(v_m-w_k+x_{p+1})}{\theta_2(x_{p+1})}
f(\bw_k,w_k) f(v_m,\bw_k)K^{p +1}_{m-1}(\bv_m|\bw_k).
\end{equation}
\end{cor}

\textsl{Proof.} Equations \eqref{rec-001} and \eqref{rec-001a} follow from the initial condition \eqref{rec-002} and the identity
\eqref{recur-2} respectively at $m_1=m-1$ and $m_1=1$.

Recursions \eqref{rec-001} and \eqref{rec-001a} also allow us to express $K_m$ in terms of $K_{m-1}$ for specific values of $v_m$. Indeed, setting
$v_m=w_m$ in \eqref{rec-001} and using $\Res g(z,w)\bigr|_{z=w}=\theta_1(\eta)/\theta'_1(0)$ we obtain
\begin{equation}\label{rec-001-1}
\Res K_{m}^{{p}}(\bv|\bw)\Bigr|_{v_m=w_m}= \frac{\theta_1(\eta)}{\theta'_1(0)} f(w_m,\bw_m)
f(\bv_m,w_m)K_{m-1}^{{p}}(\bv_m|\bw_m).
\end{equation}
Setting
$v_m=w_m-\eta$ in \eqref{rec-001a} and using $f(z-\eta,z)=0$ we obtain
\begin{equation}\label{rec-002-1}
K_{m}^{{p}}(\bv|\bw)\Bigr|_{v_m=w_m}= - \frac{\theta_2(x_{p})}{\theta_2(x_{p+1})}K^{p +1}_{m-1}(\bv_m|\bw_m).
\end{equation}

The initial condition \eqref{rec-002} and recursions \eqref{rec-001}--\eqref{rec-002-1}
correspond to the domain wall partition function of the 8-vertex model found in \cite{PRS08,YZ09,Ros09}.
Thus, similarly to models with a 6-vertex $R$-matrix, the multiple action of the upper-diagonal gauge transformed operators generates a partition function with the domain wall boundary condition.

Note that although the partition function of the 6-vertex model with domain wall boundary conditions has a determinant representation \cite{I87}, a similar result for the 8-vertex model is not yet known.
However,  a generalization of the result \cite{I87} to the  8-vertex model was obtained  in \cite{Ros09}.  It was shown that for generic $\eta$, the partition function can be presented as a sum of $2^m$ elliptic Cauchy determinants. Moreover, for  $\eta=2P/Q$ this sum reduces to only $Q/2-1$ terms for even $Q$  and $Q-1$ terms for odd $Q$. In the next section, we consider a very particular case $\eta=1/2$ that corresponds to free fermion model. We show that in this case, we obtain a single determinant as shown in \cite{Ros09}.

As a corollary to the recursion \eqref{rec-001}, one can derive an explicit form for the numerical coefficients $K_m^{p}(\bar v|\bar w)$.

\begin{cor}\label{Cor-SumPerm}
The function $K_m^p(\bar v|\bar w)$ satisfying the recursion \eqref{rec-001} and initial condition \eqref{rec-002} can be written explicitly as follows:
\begin{equation}
    K_{m}^{p}(\bar v|\bar w)=
    f(\bar v,\bar w)
    \sum_{\sigma\in S_{m}}
    \prod_{a=1}^{m}
    \left\lbrace
    \frac{%
    \theta_2(v_a-w_{\sigma(a)}+x_{\ell+r+a})%
    }{%
    h(v_a,w_{\sigma(a)})
    \theta_2(x_{\ell+r+a})}
    \prod_{k=1}^{a-1}
    \frac{%
    f(w_{\sigma(a)},w_{\sigma(k)})
    }{
    f(v_a,w_{\sigma(k)})
    }
    \right\rbrace,
    \label{cor-1}
\end{equation}
where the sum is taken over the permutations of indices.
\end{cor}

An indication for the proof for this corollary is given  in appendix~\ref{A-R}.

\subsection{Partition function in free Fermion case: A determinant representation \label{SS-FF}}

Consider the free fermion case $\eta=1/2$.
\begin{prop}\label{Prop-ff}
For $\eta=1/2$, the function $K_{m}^{p}(\bv|\bu)$ has the following explicit representation
\be{explK}
K_{m}^{p}(\bv|\bu)=\theta_2^m(0)\frac{\prod_{a>b}^m\theta_2(u_a-u_b)\theta_2(v_a-v_b)}{\prod_{a,b=1}^m\theta_1(v_a-u_b)}
\frac{\theta_2(x_{p+1}+S)}{\theta_2(x_{p+1})},
\ee
where $S=\sum_{k=1}^m(v_k-u_k)$ and $p=\ell+r$.
\end{prop}

\textsl{Proof.} We use induction in $m$. The initial condition is fulfilled. Assuming that \eqref{explK} holds for $m-1$ and using
\be{f12}
f(x,y)=\frac{\theta_2(x-y)}{\theta_1(x-y)}, \qquad \eta=1/2,
\ee
we obtain due to recursion \eqref{rec-001}
\begin{multline}\label{check}
K_{m}^{p}(\bv|\bu)=\sum_{k=1}^m \frac{\theta_2(0)}{\theta_1(v_m-u_k)}  \frac{\theta_2(x_{p+m}+v_m-u_k)} {\theta_2(x_{p+m})}
\prod_{\substack{a=1\\a\ne k}}^m \frac{\theta_2(u_k-u_a)}{\theta_1(u_k-u_a)}\prod_{b=1}^{m-1} \frac{\theta_2(v_b-u_k)}{\theta_1(v_b-u_k)}
\\
\times\theta_2^{m-1}(0)\frac{\theta_2(x_{p+1}+S-v_m+u_k)}{\theta_2(x_{p+1})}
\frac{\prod_{a>b,~a,b\ne k}^m\theta_2(u_a-u_b)\prod_{a>b}^{m-1}\theta_2(v_a-v_b)}
{\prod_{a=1}^{m-1}\prod_{b=1,~b\ne k}^m\theta_1(v_a-u_b)}.
\end{multline}
Extracting all $k$-independent factors we present \eqref{check} in the form
\be{KCtK}
K_{m}^{p}(\bv|\bu)=C_m^{p}(\bv|\bu)\tilde K_{m}^{p}(\bv|\bu),
\ee
where
\begin{equation}\label{Cm}
C_{m}^{p}(\bv|\bu)=-\frac{\theta_2^{m}(0)\prod_{a>b}^m\theta_2(u_a-u_b)\prod_{a>b}^{m-1}\theta_2(v_a-v_b)}
{\theta_2(x_{p+m})\theta_2(x_{p+1})\prod_{a=1}^{m-1}\prod_{b=1}^{m}\theta_1(v_a-u_b)},
\end{equation}
and
\begin{equation}\label{tcheck}
\tilde K_{m}^{p}(\bv|\bu)=\sum_{k=1}^m \frac{\theta_2(u_k-v_m-x_{p+m})\theta_2(u_k-v_m+x_{p+1}+S)\prod_{a=1}^{m-1}\theta_2(u_k-v_a)}
 {\theta_1(u_k-v_m)\prod_{a=1,~a\ne k}^m \theta_1(u_k-u_a)}.
\end{equation}
The sum over $k$ in \eqref{tcheck} is calculated via a standard contour integral method. Let
\begin{equation}\label{tcheck-i}
J=\frac{\theta'_1(0)}{2\pi i} \oint \frac{\theta_2(z-v_m-x_{p+m})\theta_2(z-v_m+x_{p+1}+S)\prod_{a=1}^{m-1}\theta_2(z-v_a)}
 {\theta_1(z-v_m)\prod_{a=1}^m \theta_1(z-u_a)}\,\mathrm{d}z.
\end{equation}
The integral is taken along the boundary of the fundamental parallelogram. Then $J=0$ due to periodicity of the integrand (see \eqref{A-shift}). On the other hand, the integral is equal to the sum of the residues in the poles within the integration contour. The latter are at $z=u_k$, $k=1,\dots,m$, and $z=v_m$. The sum of the residues at $z=u_k$ gives
$\tilde K_{m}^{p}(\bv|\bu)$ \eqref{tcheck}. Hence,
\be{J-res}
J=0= \tilde K_{m}^{p}(\bv|\bu)+ \frac{\theta_2(x_{p+m})\theta_2(x_{p+1}+S)\prod_{a=1}^{m-1}\theta_2(v_m-v_a)}
 {\prod_{a=1}^m \theta_1(v_m-u_a)},
 \ee
leading to
\be{J-res1}
\tilde K_{m}^{p}(\bv|\bu)=- \frac{\theta_2(x_{p+m})\theta_2(x_{p+1}+S)\prod_{a=1}^{m-1}\theta_2(v_m-v_a)}
 {\prod_{a=1}^m \theta_1(v_m-u_a)}.
 \ee
Substituting this into \eqref{KCtK} we immediately arrive at \eqref{explK}. \qed

The final result for $K_{m}^{p}(\bv|\bu)$ can be presented in the form of a determinant. For this,
we use an explicit representation for elliptic Cauchy determinant:
\be{Cauchy}
\det_m\left(\frac{\theta_1(v_j-u_k+z)}{\theta_1(v_j-u_k)}\right)=\theta_1^{m-1}(z)\theta_1(z+S)
\frac{\prod_{a>b}^m\theta_1(v_a-v_b)\theta_1(u_b-u_a)}{\prod_{a,b=1}^m\theta_1(v_a-u_b)}.
\ee
Using this formula, we can rewrite \eqref{explK} as
\begin{multline}\label{det-K}
K_{m}^{p}(\bv|\bu) = \frac{\theta_1(x_{p+1})}{\theta_1(x_{p+1}+S)}\frac{\prod_{a,b=1}^m\theta_2(v_a-u_b)}{\prod_{a>b}^m\theta_1(v_a-v_b)\theta_1(u_b-u_a)}\\
\times\det_m\left(\frac{\theta_2(0)\theta_1(2v_j-2u_k+2x_{p+1}|2\tau)}{\theta_1(x_{p+1})\theta_2(x_{p+1})\theta_1(2v_j-2u_k|2\tau)}\right).
\end{multline}
To obtain \eqref{explK} from \eqref{det-K}, one should use a particular case of \eqref{the2-the1}:
\be{t12}
\theta_1(2z|2\tau)=\frac{\theta_1(z)\theta_2(z)}{\theta_4(0|2\tau)}.
\ee

\section*{Conclusion}

In this paper, we considered the actions  of the monodromy matrix elements on the Bethe vectors in the $XYZ$ chain within the framework of the generalized algebraic
Bethe ansatz \cite{FT79}. The peculiarity of this method is that first one has to calculate the actions of the gauge transformed operators
on the pre-Bethe vectors. Knowing these actions, we can already calculate the actions of the original operators.

The actions of the monodromy matrix elements on the Bethe vectors are necessary to calculate the form factors of local operators. Indeed, if
the result of the action is expressed as a linear combination of  new Bethe vectors, then using the quantum inverse problem \cite{KitMT99,GohK00}
we reduce the form factors to scalar products. The latter were studied in \cite{SlaZZ21}. However, our calculations show that the result of
the action of any matrix element on the vector
$|\hat\Psi^{(\nu)}_n(\bu)\rangle$ generates Bethe vectors,
in which the number of parameters may differ by one from the original. In turn, in \cite{SlaZZ21}, only such scalar products were studied in which
the number of parameters in both Bethe vectors are the same. We see that such scalar products are not enough to calculate the form factors. We plan to consider scalar products of a more general form in the $XYZ$ chain in our forthcoming publications.

We have also given an example of the multiple action of gauge transformed monodromy operators on pre-Bethe vectors.
In analogy to the case of the 6-vertex $R$-matrix, we found that such multiple actions generate the partition function  of the 8-vertex model with domain wall boundary conditions $K_m^{p}(\bv|\bu)$.
We have also obtained identity \eqref{recur-2}, which is satisfied by the partition function. Note that a similar identity in models with a 6-vertex $R$-matrix plays a key role in the derivation of determinant representations for scalar products of Bethe vectors. We hope that identity \eqref{recur-2} will also be useful in the study of scalar products in the generalized algebraic Bethe ansatz.

In the particular case of free fermions, we were able to obtain an elliptic analogue of Izergin--Korepin determinant representation for $K_m^{p}(\bv|\bu)$.
We plan to continue our research in this direction in our next publication. In particular, we will give new determinant representations for the partition function in the case of rational $\eta$.

\section*{Acknowledgements}
We are grateful to A.~Zabrodin and A.~Zotov for numerous and fruitful discussions. The work of G.K. was supported by the SIMC postdoctoral grant of the Steklov
Mathematical Institute. The work of N.S. was performed at the Steklov International Mathematical Center and supported by the Ministry of Science and Higher Education of the Russian Federation (agreement no. 075-15-2022-265).

\appendix
\section{Jacobi theta-functions\label{A-JTF}}
Here we only give some basic properties of Jacobi theta-functions used in the paper. See \cite{KZ15} for more details.

The Jacobi theta-functions are defined as follows:
\be{JacTF}
\begin{aligned}
&\theta_1(u|\tau )=-i\sum_{k\in \mathbb{Z}}
(-1)^k q^{(k+\frac{1}{2})^2}e^{\pi i (2k+1)u},
\\[6pt]
&\theta_2(u|\tau )=\sum_{k\in \mathbb{Z}}
q^{(k+\frac{1}{2})^2}e^{\pi i (2k+1)u},
\\[6pt]
&\theta_3(u|\tau )=\sum_{k\in \mathbb{Z}}
q^{k^2}e^{2\pi i ku},
\\[6pt]
&\theta_4(u|\tau )=\sum_{k\in \mathbb{Z}}
(-1)^kq^{k^2}e^{2\pi i ku},
\end{aligned}
\ee
where $\tau \in \mathbb{C}$, $\Im \tau >0$, and
$q=e^{\pi i \tau}$.

To compute contour integral in section~\ref{SS-FF} we use the following shift properties:
\be{A-shift}
\begin{aligned}
&\theta_1(u+1/2|\tau)=\theta_2(u|\tau),&\qquad &\theta_2(u+1/2|\tau)=-\theta_1(u|\tau),\\
&\theta_1(u+1|\tau)=-\theta_1(u|\tau),&\qquad &\theta_2(u+1|\tau)=-\theta_2(u|\tau),\\
&\theta_1(u+\tau|\tau)=-e^{-\pi i(2u+\tau)}\theta_1(u|\tau),&\qquad &\theta_2(u+\tau|\tau)=e^{-\pi i(2u+\tau)}\theta_2(u|\tau).
\end{aligned}
\ee
To calculate matrix $\mathbf{W}^{(\ell,r)}(u)$ \eqref{Wmat0} we use the following relations:
\be{the2-the1}
\begin{aligned}
&2\theta_1(u+v|2\tau)\theta_1(u-v|2\tau)=\theta_4(u|\tau)\theta_3(v|\tau)-\theta_3(u|\tau)\theta_4(v|\tau),\\
&2\theta_4(u+v|2\tau)\theta_4(u-v|2\tau)=\theta_4(u|\tau)\theta_3(v|\tau)+\theta_3(u|\tau)\theta_4(v|\tau),\\
&2\theta_1(u+v|2\tau)\theta_4(u-v|2\tau)=\theta_1(u|\tau)\theta_2(v|\tau)+\theta_2(u|\tau)\theta_1(v|\tau).
\end{aligned}
\ee
%


\section{Identity for partition function \label{A-R}}

To prove proposition~\ref{Prop-Km}, we present $\mathbb{A}^\ell_{m,n-r}(\bv)$ as
\be{bbA-bbA}
\mathbb{A}^\ell_{m,n-r}(\bv)= \mathbb{A}^{\ell+m_1}_{m-m_1,n-r}(\bv_{\st})\mathbb{A}^\ell_{m_1,n-r}(\bv_{\so}),
\ee
where $1\le m_1<m$, and
\be{AA}
\begin{aligned}
&\mathbb{A}^\ell_{m_1,n-r}(\bv_{\so})=A_{\ell+m_1-1-r,\ell+m_1-1+r}(v_{m_1}) \cdots A_{\ell-r,\ell+r}(v_1),\\
&\mathbb{A}^{\ell+m_1}_{m-m_1,n-r}(\bv_{\st})=A_{\ell+m-1-r,\ell+m-1+r}(v_m)\cdots A_{\ell+m_1-r,\ell+m_1+r}(v_{m_1+1}).
\end{aligned}
\ee
Due to the symmetry of $\mathbb{A}^\ell_{m,n-r}(\bv)$, we consider $\bv_{\so}=\{v_1,\dots v_{m_1}\}$,  $\bv_{\st}=\{v_{m_1+1},\dots v_{m}\}$ without loss of generality.

Acting with $\mathbb{A}^\ell_{m_1,n-r}(\bv_{\so})$ on $|\psi^{\ell}_{n-r}(\bar u)\rangle$ we obtain
\begin{multline}\label{singactA-m-1}
\mathbb{A}^\ell_{m,n-r}(\bv)|\psi^{\ell}_{n-r}(\bar u)\rangle=\sum_{\{\bro_{\so},\bro_{\st}\}\vdash\{\bv_{\so},\bu\}}\frac{a(\bro_{\so})
K^p_{m_1}(\bv_{\so}|\bro_{\so})}
{f(\bv_{\so},\bro_{\so})}f(\bro_{\st},\bro_{\so})\\
\times \mathbb{A}^{\ell+m_1}_{m-m_1,n-r}(\bv_{\st}) |\psi^{\ell+m_1}_{n-r}(\bro_{\st})\rangle.
\end{multline}
Acting with $\mathbb{A}^{\ell+m_1}_{m-m_1,n-r}(\bv_{\st})$ on $|\psi^{\ell+m_1}_{n-r}(\bro_{\st})\rangle$ we obtain
\begin{multline}\label{singactA-m-2}
\mathbb{A}^\ell_{m,n-r}(\bv)|\psi^{\ell}_{n-r}(\bar u)\rangle=\sum_{\{\bro_{\so},\bro_{\st}\}\vdash\{\bv_{\so},\bu\}}\frac{a(\bro_{\so})
K^p_{m_1}(\bv_{\so}|\bro_{\so})}
{f(\bv_{\so},\bro_{\so})}f(\bro_{\st},\bro_{\so})\\
\times \sum_{\{\bro_{\sth},\bro_{\stf}\}\vdash\{\bv_{\st},\bro_{\st}\}}\frac{a(\bro_{\sth})
K^{p+m_1}_{m-m_1}(\bv_{\st}|\bro_{\sth})}
{f(\bv_{\st},\bro_{\sth})}f(\bro_{\stf},\bro_{\sth}) |\psi^{\ell+m}_{n-r}(\bro_{\stf})\rangle.
\end{multline}
The sum is taken over partitions in two steps. First, we divide the union $\{\bv_{\so},\bu\}$ into subsets $\bro_{\so}$ and $\bro_{\st}$ such that $\#\bro_{\so}=m_1$. Then we form a union $\{\bv_{\st},\bro_{\st}\}$ and divide it into subsets $\bro_{\sth}$ and $\bro_{\stf}$ such that $\#\bro_{\sth}=m-m_1$. Thus, we can say that eventually the sum is taken over partitions of the union $\{\bv,\bu\}$ into three subsets $\bro_{\so}$, $\bro_{\sth}$, and $\bro_{\stf}$ such that $\#\bro_{\so}=m_1$, $\#\bro_{\sth}=m-m_1$, and
$\bv_{\st}\cap\bro_{\so}=\emptyset$.

We can get rid of the intermediate subset $\bro_{\st}$. Since  $\{\bv_{\st},\bro_{\st}\}=\{\bro_{\sth},\bro_{\stf}\}$, we have
\be{subs}
f(\bro_{\st},\bro_{\so})=\frac{f(\bro_{\sth},\bro_{\so})f(\bro_{\stf},\bro_{\so})}{f(\bv_{\st},\bro_{\so})}.
\ee
Observe that making replacement \eqref{subs} we automatically take into account the restriction $\bv_{\st}\cap\bro_{\so}=\emptyset$, because $1/f(\bv_{\st},\bro_{\so})=0$ if there exists $v_j$ such that $v_j\in\bro_{\so}$ and $v_j\in\bv_{\st}$. Thus, we arrive at
\begin{multline}\label{singactA-m-3}
\mathbb{A}^\ell_{m,n-r}(\bv)|\psi^{\ell}_{n-r}(\bar u)\rangle=\sum_{\{\bro_{\so},\bro_{\sth},\bro_{\stf}\}\vdash\{\bv,\bu\}}
a(\bro_{\so})a(\bro_{\sth})\frac{f(\bro_{\sth},\bro_{\so})f(\bro_{\stf},\bro_{\so})f(\bro_{\stf},\bro_{\sth})}{f(\bv,\bro_{\so})f(\bv_{\st},\bro_{\sth})}
\\
\times K^p_{m_1}(\bv_{\so}|\bro_{\so})K^{p +m_1}_{m-m_1}(\bv_{\st}|\bro_{\sth})
|\psi^{\ell+m}_{n-r}(\bro_{\stf})\rangle.
\end{multline}

Let $\{\bro_{\sth},\bro_{\so}\}=\bro_0$. Then \eqref{singactA-m-3} takes the form
\
\begin{multline}\label{singactA-m-4}
\mathbb{A}^\ell_{m,n-r}(\bv)|\psi^{\ell}_{n-r}(\bar u)\rangle=\sum_{\{\bro_0,\bro_{\stf}\}\vdash\{\bv,\bu\}}
a(\bro_0)\frac{f(\bro_{\stf},\bro_0)}{f(\bv,\bro_0)} |\psi^{\ell+m}_{n-r}(\bro_{\stf})\rangle
\\
\times \sum_{\{\bro_{\so},\bro_{\sth},\}\vdash \bro_0} K^p_{m_1}(\bv_{\so}|\bro_{\so})K^{p +m_1}_{m-m_1}(\bv_{\st}|\bro_{\sth})f(\bro_{\sth},\bro_{\so})
f(\bv_{\so},\bro_{\sth}).
\end{multline}
Thus, the sum in the second line should give us $K_{m}^{p}(\bv|\bro_0)$:
\begin{equation}\label{rec-1}
 \sum_{\{\bro_{\so},\bro_{\sth}\}\vdash \bro_0} K^p_{m_1}(\bv_{\so}|\bro_{\so})K^{p +m_1}_{m-m_1}(\bv_{\st}|\bro_{\sth})f(\bro_{\sth},\bro_{\so})
f(\bv_{\so},\bro_{\sth})=K_{m}^{p}(\bv|\bro_0).
\end{equation}
We now replace $\bro_0$ with $\bw=\{w_1,\dots,w_m\}$ and set  $\bro_{\so}=\bw_{\so}$, $\bro_{\sth}=\bw_{\st}$. Then we immediately arrive at
\eqref{recur-2}.

We now turn to the proof of Corollary~\ref{Cor-SumPerm}. We use induction in $m$.
Clearly the initial condition \eqref{rec-001} can  be written in the form \eqref{cor-1} for $m=1$. To see the $m$\textsuperscript{th} iteration, let us first rewrite \eqref{rec-001} using the substitution
\begin{equation}
    f(\bar v_m, w_k)=
    \frac{f(\bar v, \bar w)}{f(\bar v_m, \bar w_k)}
    \frac{1}{f(v_m,\bar w_k)}
    \frac{1}{g(v_m,w_k)h(v_m,w_k)}.
\end{equation}
Provided \eqref{cor-1} holds for $m'<m$, we can write
\begin{multline}
    K^p_m(\bar v,\bar w)=
    f(\bar v,\bar w)
    \sum_{k=1}^{m}
    \Bigg[
    \frac{\theta_2(v_m-w_k+x_{p+m})}{h(v_m,w_k)\theta_2(x_p)}
    \frac{f(w_k,\bar w_k)}{f(v_m,\bar w_k)}
    \\
    \times
    \underset{\sigma'(m)=k}{\sum_{\sigma'\in S_{m}}}
    \prod_{a=1}^{m-1}
    \left\lbrace
    \frac{%
    \theta_2(v_a-w_{\sigma'(a)}+x_{\ell+r+a})%
    }{%
    h(v_a,w_{\sigma'(a)})
    \theta_2(x_{\ell+r+a})}
    \prod_{k=1}^{a-1}
    \frac{%
    f(w_{\sigma'(a)},w_{\sigma'(k)})
    }{
    f(v_a,w_{\sigma'(k)})
    }
    \right\rbrace
    \Bigg].
\end{multline}
This can be combined to obtain a single sum over all permutations $\sigma\in S_m$, hence it proves \eqref{cor-1}.

\end{document}